\documentclass{article}
\usepackage{spconf,amsmath,graphicx}
\usepackage{booktabs}
\usepackage{microtype}
\usepackage{subfigure}
\usepackage{gensymb}
\usepackage{multirow}
\usepackage{xcolor}
\usepackage{makecell}
\usepackage[section]{placeins}
\usepackage{textcomp}
\usepackage{url}

\renewcommand{\paragraph}[1]{
     \noindent\textbf{#1:} 
 }
 \newcommand\blfootnote[1]{%
  \begingroup
  \renewcommand\thefootnote{}\footnote{#1}%
  \addtocounter{footnote}{-1}%
  \endgroup
}

\graphicspath{ {figures/} }

\title{Low-frequency compensated synthetic impulse responses\\for improved far-field speech recognition}
\name{Zhenyu Tang \qquad Hsien-Yu Meng \quad Dinesh Manocha}

			\address{University of Maryland, College Park, MD 20742, United States \\
			    \{zhy, mengxy19, dm\}@cs.umd.edu}
\begin{document}
%
\maketitle
\begin{abstract}
We propose a method for generating low-frequency compensated synthetic impulse responses that improve the performance of far-field speech recognition systems trained on artificially augmented datasets. We design linear-phase filters that adapt the simulated impulse responses to equalization distributions corresponding to real-world captured impulse responses. Our filtered synthetic impulse responses are then used to augment clean speech data from LibriSpeech dataset~\cite{panayotov2015librispeech}. We evaluate the performance of our method on the real-world LibriSpeech test set. In practice, our low-frequency compensated synthetic dataset can reduce the word-error-rate by up to 8.8\% for far-field speech recognition.


\end{abstract}

\section{Introduction}
Automatic speech recognition (ASR) has been an active area of research. In recent years, there has been considerable improvement in accuracy due to the use of learning-based techniques. Deep neural networks (DNNs) are proven to be a powerful tool for fitting a target data distribution~\cite{yu2017recent}, given enough training examples. However, it is also known that DNNs are generally not robust to out-of-domain test data~\cite{yoshioka2012making}. To overcome this issue, there is increasing interest in collecting larger, more diverse datasets in various research communities such as computer vision, natural language processing, and speech understanding. Further, thanks to the availability of large labeled recorded speech corpora~\footnote{\url{https://www.openslr.org/resources.php}}, state-of-the-art ASR systems are highly accurate in specific conditions and are beginning to outperform humans on public benchmarks~\cite{amodei2016deep}. 
\blfootnote{This work is supported in part by ARO grant W911NF-18-1-0313, NSF grant \#1910940, Adobe, Facebook and Intel.}

Near-field (close-talking) speech refers to a situation in which the talking person is in close proximity to the listener/microphone, and there are no strong reverberation effects in the speech signal. Near-field speech is almost anechoic and has a high signal-to-noise ratio (SNR), so it is also referred to as ``clean" speech. The opposite of near-field speech is far-field (distant-talking) speech, which means there is a non-negligible distance between the speaker and the listener and strong room reverberation effects can be present. As a result, far-field speech can be severely affected by room reverberation and environmental noise and exhibits significantly different statistical properties~\cite{yoshioka2012making}. A DNN model trained using clean speech is unlikely to perform well for far-field speech~\cite{ko2017study}. The challenge for far-field ASR is that large possible variations in the reverberation patterns are associated with the room layout and speaker-listener setup. This makes it challenging to gather or create a large enough far-field ASR training set that works for all situations. Therefore, using data augmentation to enrich existing speech data and approximate far-field speech becomes the most viable solution~\cite{ko2017study,karafiat2017training,ravanelli2017realistic,szoke2019building}.

Far-field speech can be effectively simulated by convolving clean speech with an impulse response (IR), and adding ambient noise at different SNRs. One difficulty during augmentation is obtaining IRs that characterize various acoustic environments. Real-world IRs can be recorded using sine-sweeps~\cite{farina2000simultaneous}, but this is a laborious and intrusive process and requires special hardware and human expertise. An alternative is to use simulated IRs for augmentation. Many IR simulators have been designed to calculate room reverberations according to user defined acoustic environments~\cite{savioja2015overview}. One caveat of the most widely used IR simulators is that they are unable to model all acoustic effects present in the real-world, specifically, low-frequency wave acoustic effects. Consequently, techniques developed for IR simulation are based on geometric acoustics and are less accurate for low-frequency sound, which can be crucial to speech understanding: low-frequency components in speech has been proven to improve speech intelligibility~\cite{brown2009low} and contribute strongly to tone recognition~\cite{luo2006contribution}. An analysis into the filters learned by a convolutional neural network reveals the convolution layer learns to emphasize low-frequency speech regions, which cover the fundamental frequency of human speech~\cite{muckenhirn2018towards}. Lacking accuracy in low-frequency acoustic simulation can potentially degrade the performance of ASR systems.



\paragraph{Main Results} 
To overcome the limitation of current IR simulations, we propose a method to generate realistic synthetic data with low-frequency compensation for training far-field ASR systems. Our main contributions include:
\begin{itemize}
    \item A generative model for real-world room equalization (EQ), which accounts for room-related frequency gains of sound, including low-frequency wave effects.
    \item A scheme to generate low-frequency compensated IRs using ray-tracing based acoustic simulation for far-field ASR training with synthetic dataset.
    \item Far-field ASR evaluations based on real-world LibriSpeech~\cite{panayotov2015librispeech} test set, which show our synthetic dataset with compensation outperforms the non-compensated dataset by up to 8.8\%.
\end{itemize}

The rest of the paper is organized as follows: In Section 2 we introduce IR simulation techniques and related work in ASR that uses synthetic IRs. Section 3 gives details of our novel EQ compensation method. Section 4 describes the far-field ASR evaluation setup and presents the results. We conclude the paper in Section 5. Data and code from this paper are released for reproducibility~\footnote{Project website \url{https://gamma.umd.edu/pro/speech/eq}}.

\section{Related Work}
Compared with the large number of labeled clean speech datasets, the scarcity of labeled far-field speech data is a limiting factor for training far-field ASR systems. However, far-field speech data can be artificially created by convolving clean speech with an IR, then adding environmental noise, which is one speech augmentation method. Far-field speech augmentation conventionally uses the equation~\cite{ko2017study}:
\begin{equation}
\label{eq:augment}
x_{r}[t]=x[t] * h_{s}[t]+\sum_{i} n_{i}[t] * h_{i}[t]+d[t],
\end{equation}
where $*$ denotes a linear convolution, $x[t]$ represents the clean speech signal, $n_i[t]$ represents the point-source noise, $h_s[t]$ and $h_i[t]$ are IRs corresponding to the signal and noise sources, respectively, $d[t]$ represents any other ambient noise, and $x_r[t]$ represents the augmented far-field speech signal that is used for training. The IR essentially describes how sound energy is reflected and scattered by all surfaces in an acoustic environment, given the sound source and listener locations~\cite{kuttruff2016room}. Many physically-based sound simulation methods have been developed over the past few decades for computations of IRs in synthetic scenes~\cite{funkhouser2003survey,savioja2015overview}. They are able to compute the IR for a given source and listener location, based on user specified scene geometry and acoustic material properties. Among these methods, wave-based methods~\cite{sakamoto2006numerical,raghuvanshi2009efficient} are the most accurate, but are computationally expensive and cannot scale with high simulation frequency and scene complexity. In practice, the image method~\cite{allen1979image} is widely used due to its efficiency, but it only models high frequency specular reflections. Recent studies~\cite{tang2019regression,tang2019improving} using a more accurate ray-tracing based geometric acoustic simulation have shown its advantages over the image method because it also models diffuse reflections that are more important for late reveberation effects. The limitation of both methods is their assumption that sound waves propagate as geometric rays~\cite{schissler2017interactive,schissler2018interactive}, which holds for high-frequency waves whose wavelengths are significantly smaller than scene primitives. However, this assumption does not hold for low-frequency waves. Therefore, these methods will not work well for low-frequency components of the sound, which inevitably introduces significant simulation error at low frequencies under 500Hz compared to accurate wave-based solvers~\cite{morales2018optimizing}.

Far-field ASR experiments that aim to compare the effectiveness of using simulated IRs against real IRs confirm that real IRs are superior in training better ASR systems~\cite{szoke2019building}. The main drawback of geometric acoustic simulation is the absence of low-frequency wave effects such as diffraction~\cite{torres2001computation} and room resonance~\cite{avis2000active,tang2019scene}, of which sound diffraction is a less noticeable phenomenon. Room resonance creates wave modes that either boost or diminish the frequency response at different frequency bands. While the frequency response of an IR simulated using geometric methods is mostly flat due to the ray assumption, the response with room modes has visible irregularities, especially at low-frequencies where room resonance is most dominant. Room resonance is normally undesirable for sound field reproduction and various methods have been devised to remove this effect~\cite{cecchi2018room}. However, in this paper, we aim at compensating the missing room resonance by room EQ compensation, which helps reducing the performance gap between using simulated and real IRs for far-field ASR training.

\section{Proposed Method}
Our method uses real-world IR as a reference to guide our synthetic IR compensation. The analysis and the matching of sub-band room EQs are essential stages before designing our compensation filters.

\subsection{Equalization Analysis}
\begin{figure}[htbp]
    \centering
    \includegraphics[width=0.9\linewidth]{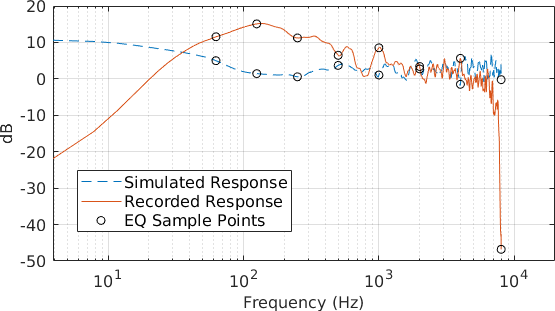}
    \vspace{-1em}
    \caption{Recorded and simulated frequency responses (magnitude) and EQ sample points with logarithmic scale for x-axis. The recorded frequency response has more variation at certain frequency ranges. Our EQ sample points have higher density over low frequencies, where the simulated and recorded responses are most distinct.}
    \label{fig:freq_response}
    \vspace{-1.5em}
\end{figure}

In this paper, without loss of generality, we work with audio at a 16kHz sample rate, and any other sample rates that are compatible with the speech datasets (e.g. LibriSpeech) can also be used. IRs are converted to frequency responses using Fourier transform with a window size of 512. An example showing a simulated frequency response and a recorded frequency response is shown in Figure~\ref{fig:freq_response}, in which the range of frequency gains of the simulated response is relatively flat compared with those of the recorded response. Also, there are frequency regions that are significantly amplified in the low-frequency part (below 1000Hz) of the recorded response, which covers typical frequencies of low room modes that are dominant~\cite{avis2000active}.


\subsection{Equalization Matching}
To bridge the gap between simulated and recorded responses, we first analyze the statistical properties of real-world frequency responses. Instead of fully representing the response curve using thousands of discrete samples, we choose 8 frequency sample points at [62.5, 125, 250, 500, 1000, 2000, 4000, 8000] Hz for each curve, representing the sub-band EQ gains of a frequency response and annotate them in Figure~\ref{fig:freq_response}. The actual degree of freedom of this representation is 7 because signal gains are relative and we offset all sub-band EQ gains by using the gain at 1000Hz as a reference. While we use 8 frequency sample points, a similar analysis can be performed with different number of sample points.

\begin{figure}[htbp]
    \centering
    \includegraphics[width=1.0\linewidth]{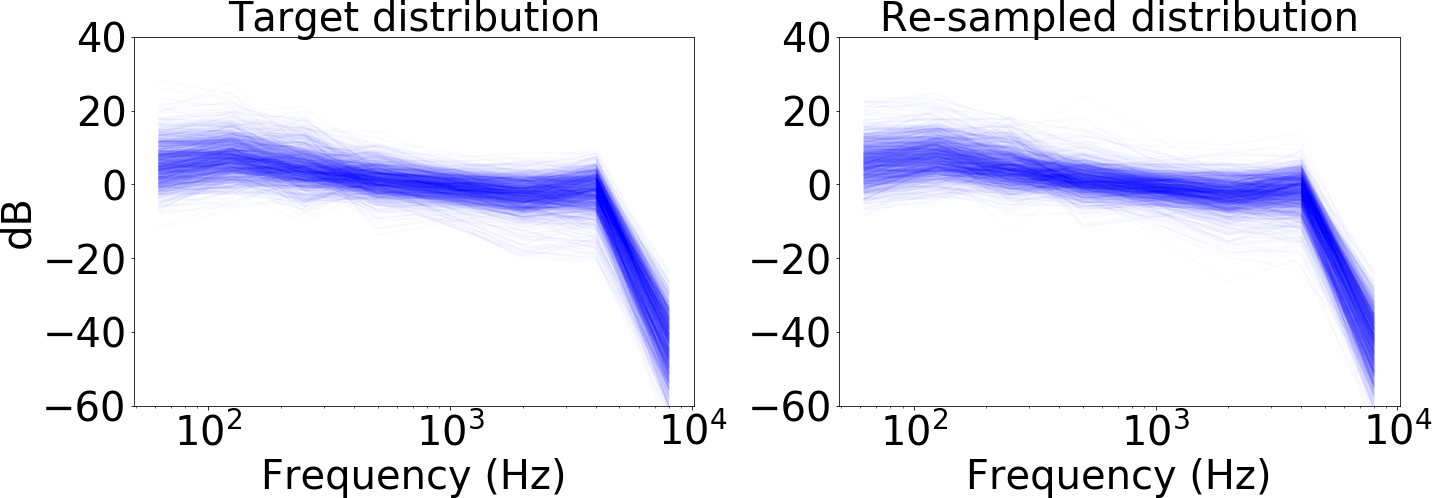}
    \vspace{-2em}
    \caption{Target sub-band EQ distribution (left) and re-sampled distribution (right) with fitted GMM. There is a good match between the two distributions, meaning that our GMM has successfully captured the variation of the target distribution.}
    \label{fig:eq}
    \vspace{-1em}
\end{figure}

Based on this representation, we calculate and extract sub-band EQs for a set of recorded IRs, collected from the BUT Reverb Database (ReverbDB)~\cite{szoke2019building}. The BUT ReverbDB contains 1891 IRs and 9114 background noises (both with some repetitions), recorded in 9 different real-world environments. The extracted sub-band EQ curves of IRs are visualized in Figure~\ref{fig:eq} (left). We see that each sub-band follows different mean and standard deviations, therefore, we use a Gaussian mixture model (GMM). We choose 7 as the number of mixture components for GMM, which is the same as the variable dimension. We then re-sample the same amount of sub-band EQs as the input and visualize them in Figure~\ref{fig:eq} (right) under the same scale. Overall, the fitted GMM closely resembles the target distribution.

\subsection{Compensation Filters}
\begin{figure}[htbp]
    \centering
    \includegraphics[width=0.8\linewidth]{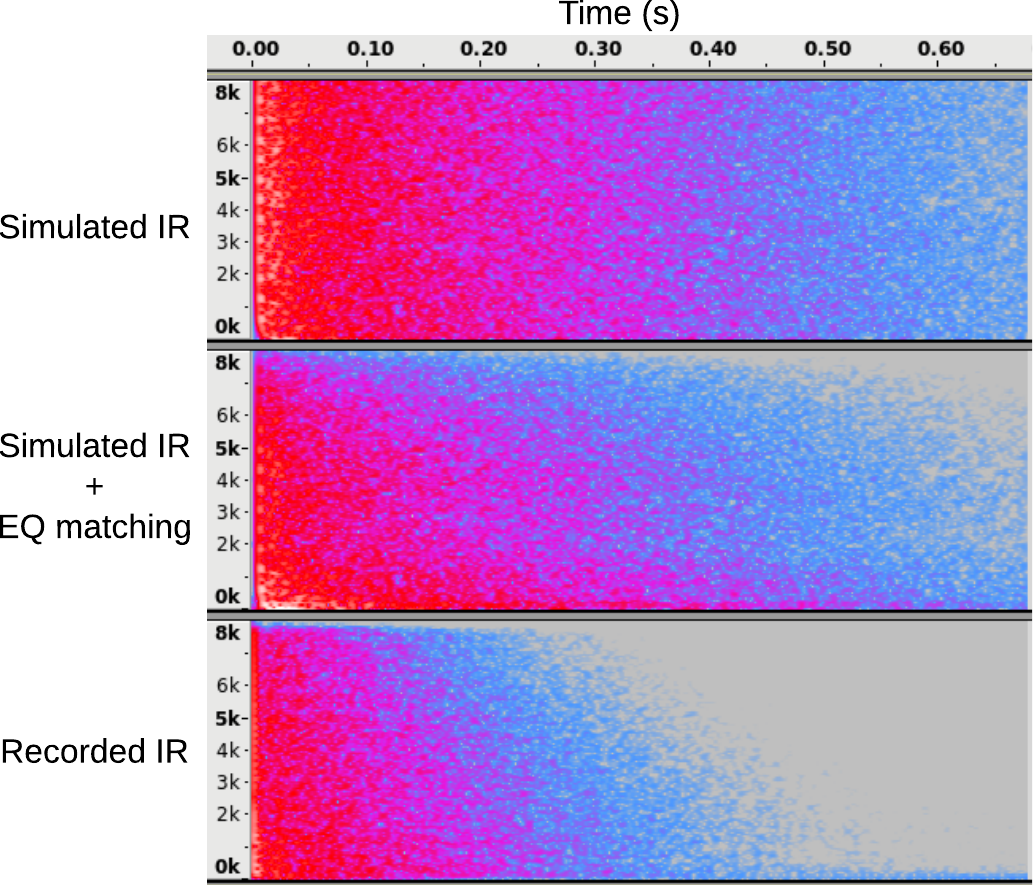}
    \vspace{-1em}
    \caption{Spectrograms of a simulated IR, the simulated IR after EQ matching, and an example recorded IR. The simulated IR after EQ matching has energy distribution over low and high frequencies (due to room EQ) more similar to a recorded IR, despite the difference between their reverberation times.}
    \label{fig:comp}
    \vspace{-1em}
\end{figure}

We present our EQ compensation filters that are used to modify synthetic IRs generated by acoustic simulators to match the real-world IRs. After computing a generative GMM for the real-world sub-band EQs, for each simulated IR, we can sample a new sub-band EQ from this distribution and set it as our target for IR compensation. We then evaluate the sub-band EQ gain (vector of length 7) of the original simulated IR and calculate the difference between the original gain and the newly sampled target sub-band EQ. This difference is the additional desired EQ gains we need to apply to the original IR. We supply the desired EQ gains and solve the time domain finite impulse response (FIR) filter coefficients using the window method~\cite{smith1997scientist} with 511 taps under our 16kHz sample rate. After the simulated IRs are filtered by these EQ compensation filters, their sound energy will be re-balanced among low-frequency and high-frequency wave components. As depicted in Figure~\ref{fig:comp}, the spectrogram of a compensated IR has similar response pattern among low-frequency and high-frequency components like recorded IRs. Note that we do not match recorded IRs exactly pair-by-pair, which would correspond to duplicating the real data, but we aim to generate new simulated data that has frequency domain energy distributions similar to the real IRs.

\label{sec:filters}
\section{Experiments and Results}
\label{sec:results}
We evaluate our method by training and testing on public ASR datasets under equivalent conditions and report the word-error-rate (WER) of each trained system. The main comparison is between simulated IRs with and without our low-frequency EQ compensation. In addition, we also train a baseline model using clean speech and an oracle mode using real IRs with the same distribution as the test set.

\begin{table}[htbp]
\centering
\vspace{-1em}
\caption{Augmented dataset overview. We use *.real.rvb to denote dataset augmented using real IRs, *.sim.rvb to denote dataset augmented using simulated IRs, and *.sim.EQ.rvb to denote dataset augmented using EQ compensated simulated IRs. The training sets (train.*) and development sets (dev.*) are augmented based on either simulated or recorded IRs, and the test set (test.*) is based on only real IRs and is used to evaluate all trained models in this paper. Durations of each augmented dataset remain the same as the original LibriSpeech dataset.}
\label{tab:dataset}
\begin{tabular}{llll}
\toprule
Augmented dataset    & Hours & \#IRs & Source in LibriSpeech \\\hline
train.real.rvb  & 460   & 773  & train-clean-\{100,360\} \\
dev.real.rvb   & 5.4   & 194  & dev-clean \\
test.real.rvb   & 5.4   & 242  & test-clean\\\hline
train.sim.rvb   & 460   & 773  & train-clean-\{100,360\} \\
dev.sim.rvb   & 5.4   & 194  & dev-clean \\\hline
train.sim.EQ.rvb & 460   & 773  & train-clean-\{100,360\} \\
dev.sim.EQ.rvb   & 5.4   & 194 & dev-clean  \\\bottomrule
\end{tabular}
\vspace{-1.5em}
\end{table}

\begin{table*}[!htbp]
\centering
\caption{Far-field ASR results with far-field LibriSpeech test set. Note that in each row, the development set has the same distribution as the training set, and there is a mismatch between the development set and the test set except for train.real.rvb -- our oracle model. WER is reported for tri-gram phone (tg\{large,med,small\}) and four-gram phone (fglarge) language models, and online decoding using tgsmall. Our proposed EQ compensated training set (train.sim.EQ.rvb) consistently outperforms the simulated training set without compensation (train.sim.rvb) on all test conditions, while having a much smaller gap between dev and test WERs. Best test results using synthetic datasets are marked in \textbf{bold}.}
\label{tab:results}
\begin{tabular}{@{\extracolsep{4pt}}ccccccccccc@{}}
\toprule
\multirow{2}{*}{Training data}      & \multicolumn{5}{c}{dev WER {[}\%{]}}                      & \multicolumn{5}{c}{test WER {[}\%{]}}                                                           \\\cline{2-6}\cline{7-11}
                              & fglarge & tglarge & tgmed & tgsmall & online & fglarge        & tglarge        & tgmed          & tgsmall        & online         \\\hline
train.clean (Baseline)              & 3.56    & 3.60    & 4.44  & 4.97    & 4.96   & 73.97          & 74.15          & 75.54          & 76.08          & 76.29          \\
train.real.rvb (Oracle)               & 11.94   & 12.58   & 15.24 & 16.81   & 16.82  & 11.80          & 12.42          & 15.09          & 16.32          & 16.32          \\\hline
train.sim.rvb                 & 11.19   & 11.98   & 14.52 & 15.91   & 15.89  & 15.56          & 16.27          & 19.19          & 20.84          & 20.76          \\
train.sim.EQ.rvb              & 14.70   & 15.53   & 18.64 & 20.45   & 20.42  & \textbf{14.19} & \textbf{14.92} & \textbf{17.91} & \textbf{19.66} & \textbf{19.67} \\ \bottomrule
\end{tabular}
\vspace{-1em}
\end{table*}

\subsection{Data Preparation}
\subsubsection{Benchmark}
Our goal is to train far-field ASR systems. While there exists far-field datasets and tasks including the AMI task~\cite{hain2006ami}, the REVERB challenge~\cite{kinoshita2013reverb}, the ASpIRE task~\cite{harper2015automatic}, the CHiME-4 challenge~\cite{vincent2017analysis}, etc., few real-world IRs are generally provided, especially for training purposes. These do not supply sufficient information about the recording condition for test sets for us to run controlled comparisons in terms of data augmentation. 325 real-world IRs are collected from different sources in~\cite{ko2017study} for far-field ASR tests, however, these IRs are still not guaranteed to match the test condition in~\cite{ko2017study}.

To overcome this difficulty, we create a new test set using clean speech from LibriSpeech~\cite{panayotov2015librispeech} and real-world IRs and noises from BUT ReverbDB~\cite{szoke2019building}. In~\cite{szoke2019building}, there is almost no performance gap found between using far-field speech created by mixing these real-world recordings and physically re-transmitting the signal in various environments. Therefore, our test condition is still practical, with known IRs and information of their corresponding acoustic environments. After eliminating repeated recordings and uncommon condition recordings that are difficult to simulate (e.g. stairways), we extract 1209 recorded IRs from BUT ReverbDB, which are randomly split into subsets of \{773, 194, 242\} IRs for training, development, and testing purposes, respectively. Detailed composition is in Table~\ref{tab:dataset}.

\subsubsection{Geometric Acoustic Simulation}
One advantage of using BUT ReverbDB is the detailed meta-info accompanying each recording. For each recording, the environment size (ranging from $100m^3$ to $2000m^3$) and dimensions, loudspeaker location, and microphone location are well documented. We use this as input to a state-of-the-art geometric acoustic simulator~\cite{tang2019improving,schissler2017interactive} which simulates both specular and diffuse reflections up to 200th order; we generate the same number of simulated IRs as recorded IRs from BUT ReverbDB. However, a one-to-one mapping from simulated IRs to recorded IRs should not be expected, because the reverberation time ($T_{60}$) still depends on room materials. These material properties are frequency-dependent and difficult to capture. As a result, we randomly sample the room materials to obtain IRs that have $T_{60}$ uniformly distributed between 0.2s and 2.0s, which accounts for most real-world reverberation conditions. In addition to this set of simulated IRs, we use the filters described in Section~\ref{sec:filters} for EQ and create another set of compensated IRs.


\subsubsection{Far-field Speech Augmentation}

Overall, we follow the common augmentation procedure according to Equation~(\ref{eq:augment}) to mix either recorded or simulated IRs with clean speech, and add ambient noise. An additional step in our pipeline is to shift the convolved signal such that it has the same start as the original clean speech. This is essential because misalignment due to convolution can directly result in WER increases. We perform this alignment by detecting the direct response in IRs and advance the convolved signal by this time amount. Clean speech from LibriSpeech is used, specifically the train-clean-360, train-clean-100, dev-clean, and test-clean sets. Note that we do not use the train-other-500 or test-other set of LibriSpeech, because the recording quality of them is considered to be lower than ``clean" sets, which may bias our results. 

Recorded IRs, simulated IRs, and compensated IRs are pre-split into disjoint subsets containing 773 and 194 IRs for generating far-field training and development sets, while another 242 recorded IRs are reserved for creating the test set. We randomly sample recorded noises from BUT ReverbDB for all augmented datasets and do not extensively experiment with the noise addition strategy, as that is not the focus of this work and has been studied in~\cite{ko2017study,szoke2019building}. The final augmented dataset composition and properties are shown in Table~\ref{tab:dataset}.

\subsection{Training Procedure}
We use Kaldi~\cite{Povey_ASRU2011} toolkit for our far-field ASR training and testing. We modify  its original recipe for LibriSpeech by removing the stages that utilize the lower quality ``other" datasets to use our customized augmented datasets. Time-delay neural networks (TDNNs)~\cite{peddinti2015time} are trained with the lattice-free maximum mutual information (LF-MMI) criterion~\cite{povey2016purely}, for each of our augmented training sets using a multi-GPU setup. In addition, we also include the unmodified ``clean" datasets from LibriSpeech as our baseline to observe the domain gap between clean speech and far-field speech. On a server with 32 Intel(R) Xeon(R) E5-2630 v3 CPUs and 4 Nvidia TITAN X (Pascal) GPUs, each model requires approximately 80 hours to train and test.

\subsection{Results and Analysis}
We present our far-field ASR results in Table~\ref{tab:results}. We report WER decoded using 4 language models included with Kaldi: \emph{fglarge, tglarge, tgmed, tgsmall}, where \emph{tg} refers to tri-gram phone language model (with 3 different sizes) and \emph{fg} refers to four-gram phone language model, as well as online decoding. 

First we observe that all models are much better than the baseline model, as expected. The baseline model trained on clean speech achieves the lowest dev error, but has the largest performance gap between dev error and test error. Also, its dev error is much lower than train.real.rvb, which means clean speech is easier to fit than noisy reverberant speech using DNNs.

The best WER obtained using train.sim.rvb (without EQ) is 15.56\% with \emph{fglarge} model. The corresponding WER of train.sim.EQ.rvb (proposed EQ compensated) is 14.19\%, which translates to a relative 8.8\% WER improvement. In addition, the gap between dev and test WERs of train.sim.EQ.rvb is also significantly smaller than that of train.sim.rvb, which means our synthetic data with EQ compensation has data distribution that is more similar to real data, than data without EQ compensation. 

Furthermore, the TDNN we used from Kaldi is designed and tuned for clean speech training, and we do not modify it here to keep the comparison fair. However, the dev error of train.sim.EQ.rvb is not as low as other models, meaning there is still room for its WER improvement by fine-tuning the DNN structure or hyper-parameters.

\section{Discussion and Future Work}
In this paper, we present a method to generate more realistic synthetic IRs with low-frequency compensation for far-field ASR training. We showed that by using our method, the WER of a far-field ASR task on real-world test data can be reduced by up to 8.8\%, significantly reducing the performance gap between the use of synthetic data and real data for training. One limitation of this work is that simulated IRs are not generated on a pair-by-pair basis with respect to recorded IRs, due to the lack of acoustic material information about the room, which is usually difficult to acquire. Moreover, we neither fine-tune the synthetic models using real data, nor combine synthetic and real data for training, which is a good area for future work to develop more robust far-field ASR systems.

\bibliographystyle{IEEEtran}
\bibliography{mybib}

\end{document}